\documentclass[prl,aps]{revtex4}
\usepackage{graphicx}
\usepackage{amssymb}
\usepackage{amsmath}
\usepackage{lscape}
\usepackage{textcomp}
\usepackage{epsfig}

\begin{document}

\title{Effect of topology on the critical charge in graphene}

\author{Baishali Chakraborty$^{1}$}
\email{baishali.chakraborty@saha.ac.in}
\author{Kumar S. Gupta$^{1}$}
\email{kumars.gupta@saha.ac.in}
\author{Siddhartha Sen$^{2}$}
\email{siddhartha.sen@tcd.ie}
\affiliation{$^{1}$Theory Division, Saha Institute of Nuclear Physics, 1/AF Bidhannagar, Calcutta 700064, India\\
$^{2}$Trinity College, Dublin, Ireland}

\date{\today}

\newcommand{\be}{\begin{equation}} \newcommand{\ee}{\end{equation}}
\newcommand{\bea}{\begin{eqnarray}}\newcommand{\eea}{\end{eqnarray}}
\begin{abstract}
We show that the critical charge for the Dirac excitations in gapless graphene depends on the spatial topology of the sample. In particular, for graphene cones, the effective value of the critical charge can tend towards zero for a suitable angle of the conical sample. We discuss the nature of the scattering phase shifts, quasi-bound state energies and local density of states for a gapless graphene cone and determine the dependence of these physical quantities on the sample topology. 

\end{abstract}
\maketitle
\section{Introduction}

The fabrication of monolayer graphene \cite{novo1,novo2,zhang} provides an opportunity to study the effect of sample topology on its 
electronic properties \cite{crespi1,crespi2,osi1,voz1,furtado,stone1}. The low energy excitations of the gapless graphene are described by a 
two dimensional massless Dirac equation \cite{wall,geim,rmp,mele,sem}. These excitations are negatively charged and behave like electrons 
with the Fermi velocity $v_F \approx10^6 m/s $. In graphene, even a small external charge impurity $Ze \sim 1$ can produce strong 
nonperturbative effects, since the effective Coulomb interaction strength $\alpha = \frac{Ze^2}{\hbar \kappa v_F} \sim 1$, where the 
dielectric constant $\kappa \sim 5$. Thus, graphene provides an ideal system to analyze the combined effects of topology and strong electric 
fields on its electronic properties.

The idea of a critical charge plays an interesting role in this context \cite{castro,levi1,levi2,us1,kats1}. It is well known that massless 
Dirac fermions can tunnel through large external potentials, thereby avoiding trapping or formation of bound states. This phenomenon, known 
as Klein tunneling \cite{kl1,kl2,kats2,bridge,QED}, is possible in graphene due to the chiral nature of the Dirac excitations 
\cite{kats1,kats2}. However, when the effective strength of the external charge exceeds a certain critical value $\alpha_c$, the low energy 
excitations can form quasi-bound states \cite{castro,levi1,levi2,us1,kats1}. For such states the wavefunctions oscillate rapidly near the 
position of the charge impurity, analogous to {\it zitterbewegung} in strong field QED \cite{greiner,QED}. For planar graphene, the  critical
charge $\alpha_c = \frac{1}{2}$. The existence of such quasi-bound states and their effect on transport properties and local density of states 
(LDOS) in planar graphene have attracted considerable attention \cite{castro,levi1,levi2,kats1}. 

In this paper we explore how a nontrivial topology of the sample may affect the critical charge and the corresponding strong field QED effects 
in graphene. We address this issue for graphene cones, which are obtained by introducing local defects in the hexagonal lattice structure, 
for example, in the form of a pentagonal ring \cite{crespi1,crespi2}. The conical topology manifests itself in the form of nontrivial 
holonomies for the pseudoparticle wavefunctions. Such holonomies can be modelled through the introduction of a magnetic flux tube. Thus an 
electric charge localized at the apex of the cone can be equivalently described by a suitable combination of electric charge and magnetic 
flux tube. We show that the critical charge $\alpha_c$  depends on the angle of the graphene cone. For certain angle of the graphene cone, the topological effects lead to the identification of the two Fermi points \cite{crespi1,crespi2}. This leads to the surprising conclusion that for a certain value of the angle of the cone, the critical charge is zero. We discuss the effect of the conical topology on the scattering phase shifts, quasi-bound state energies and local density of states (LDOS) in graphene. 

This paper is organized as follows. In the next Section we set up the Dirac equation for gapless graphene cone with a point charge at the 
apex. This is followed by the analysis of the corresponding spectrum, where we obtain the scattering phase shits, quasi-bound state energies 
and local density of states (LDOS) and show how these physical quantities depend explicitly on the sample topology. We also briefly comment 
on the RG flow of the strength of the charge impurity. We end this paper with some discussion and outlook. 

\newpage

\section{Dirac equation for a graphene cone with a Coulomb charge}

In this section we set up the Dirac equation for a graphene cone with a coulomb charge at the apex of the cone.  The low energy properties of the quasiparticle 
states near the Dirac points in graphene can be described by the four component Dirac wave function \cite{wall, geim, rmp} given by
\begin{eqnarray}
\label{h1.0}
\Psi= \left( 
\begin{array}{c}
 \Psi_{A} \\
 \Psi_{B} 
\end{array}
\right), ~~~{\mathrm {where}} ~~ 
\Psi_A = \left( 
\begin{array}{c}
 \Psi_{A_+} \\
 \Psi_{A_-} 
\end{array} \right)
~~~{\mathrm {and}} ~~ 
\Psi_B = \left( 
\begin{array}{c}
 \Psi_{B_+} \\
 \Psi_{B_-} 
\end{array} \right)
\end{eqnarray} 
The pseudospin indices $A$ and $B$ label the two sublattices of the primitive cell of graphene and the valley indices $+$ and $-$ label the two inequivalent Dirac points $\mathbf {K}_{+}$ and $\mathbf{K}_{-}$ respectively. 

Considering low energy excitations about the Dirac point $K_{+}$, Dirac equation for a planar gapless graphene in the presence of a Coulomb 
charge $Ze$ is given by 
\begin{equation}
\label{h1.1}
 H \Psi = \left [ -i\hbar v_F (\sigma_1 \partial_x + \sigma_2 \partial_y) + \sigma_0 \left ( \frac{ - \alpha}{r} \right ) 
\right ] \Psi = E \Psi, 
\end{equation}
where $r$ is the radial coordinate in the two dimensional $x-y$ plane. The Pauli matrices $\sigma_{1,2,3}$ and the identity matrix $\sigma_0$
act on the pseudospin indices$A,B$. The Coulomb interaction strength $\alpha = \frac{Ze^2}{\hbar \kappa v_F}$. From now on we set 
$\hbar = v_F = 1$.
In a two dimensional plane, the angular boundary condition for a  Dirac spinor as it goes around a closed path is given by
\begin{eqnarray}
\label{h1.2}
\Psi(\mathbf{r},\theta=2\pi)=e^{i\pi\sigma_3}\Psi(\mathbf{r},\theta=0).
\end{eqnarray}

A cone is obtained from the two dimensional plane by introducing a topological defect, which modifies the angular boundary condition (\ref{h1.2}). Following \cite{crespi1},
consider a cone, formed by removing the sector $AOB$ (see Fig.1a.) and then identifying the edge $OA$, labeled by $\theta=0$ with the 
edge $OB$, labeled by $\theta=2\pi$. The sector $AOB$ subtends an angle $\frac{2\pi}{6}$ at the centre $O$. Due to this identification the frame $\{\hat{e}_x,\hat{e}_y\}$ becomes discontinuous across the joining line. This problem can be solved by choosing a new set of frames given by 
\begin{eqnarray}
\label{h1.5}
\hat{e}_{x^{\prime}} = \hat{e}_\theta~~~~~~~~~~\mbox{and}~~~~~~~~~~~\hat{e}_{y^{\prime}} = -\hat{e}_r,
\end{eqnarray}
which is rotated with respect to old frame $(\hat{e}_x,\hat{e}_y)$ by an angle $\phi =\theta + \frac{\pi}{2}$
in the counter clockwise direction, see Fig.1b. To keep the form of the Hamiltonian the same, the wave function has to be transformed by $\mbox{exp}(i\phi \sigma_3/2)$. Similarly, if $n$ sectors are removed from the plane where $n$ can take only discrete values $1,2,3,4,5$, the wave function should be transformed as 
\begin{eqnarray}
\label{h1.6}
\Psi(\mathbf{r},\theta=2\pi)=-e^{i2\pi(1-\frac{n}{6}) \frac{\sigma_3}{2}}\Psi(\mathbf{r},\theta=0).
\end{eqnarray} 
Note that an additional negative sign appears in the boundary condition since we take the angular part of the wavefunction to be $e^{ij\theta}$, $j$  being a half integer \cite{crespi1,crespi2}.

\begin{figure}
[ht] 
\centering
\includegraphics[bb=380 14 112 250]{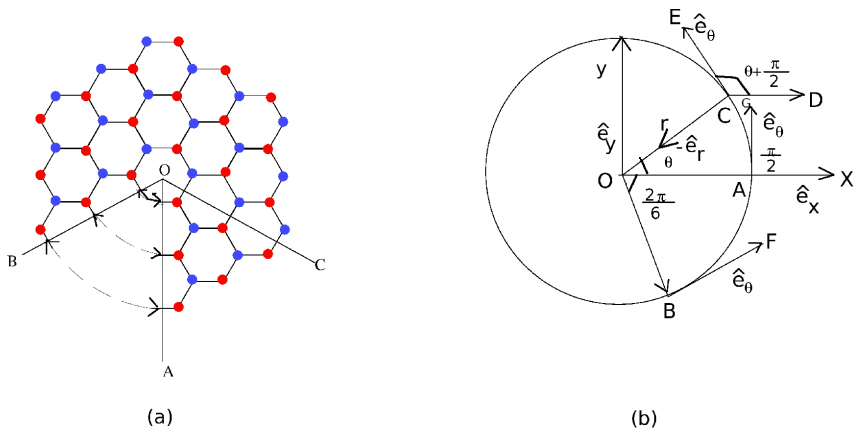}
\caption{(a)Formation of a cone from plane graphene sheet by cut and paste procedure and (b)Rotation of the coordinate frame due to its new 
         orientation. In Fig. (a) blue atoms represent sublattice A and red atoms represent sublattice B.}
\label{fig:1}
\end{figure}

\begin{figure}
[ht] 
\centering
\includegraphics[bb=180 14 12 180]{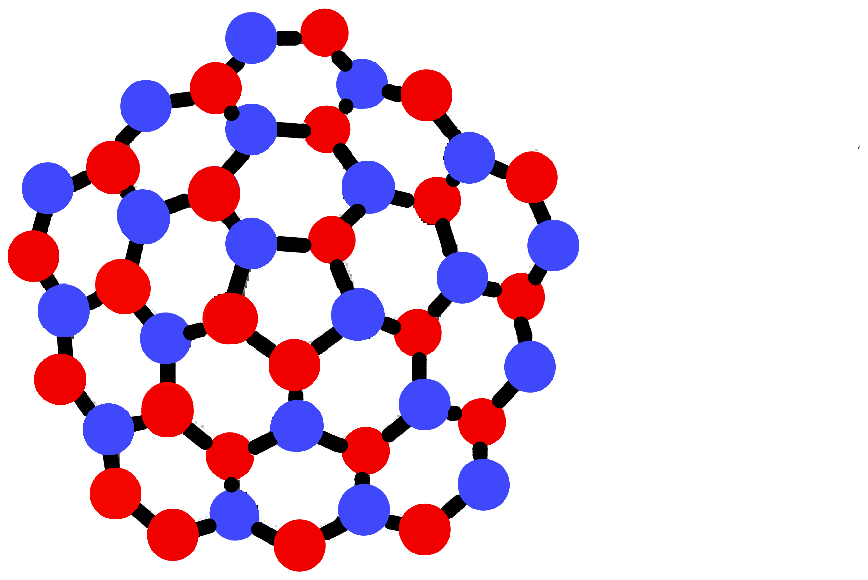}
\caption{A Graphene cone formed by removing an angular sector of $\frac{2\pi}{6}$ from the plane sheet.}
\label{fig:2}
\end{figure}
When n is odd,  an additional phase change is required due to the breaking of the bipartite nature of the hexagonal lattice.
It can be seen from the Fig.(\ref{fig:2}), that the removal of a wedge of opening angle $\frac{2\pi}{6}$ introduces a pentagonal defect. When the cone is formed by removing a single wedge of angle $\frac{2\pi}{6}$, the adjacent points on two sides of the identification line belong to the same sublattice. This feature is valid for all allowed odd $n=1,3,5$ and leads to a system with only one Fermi point. Considering all the factors, the angular boundary condition satisfied for a general value of $n$ is given by  \cite{crespi1,crespi2}.
\begin{eqnarray}
\label{h1.8}
\Psi(\mathbf{r},\theta=2\pi)=-e^{i2\pi[-\frac{n\tau_2}{4}+(1-\frac{n}{6}) \frac{\sigma_3}{2}]}\Psi(\mathbf{r},\theta=0).
\end{eqnarray}
Here $\tau_2$ acts on the valley indices $(+,-)$.
Note that for $n=0$, the angular boundary condition for the planar graphene is recovered. For the allowed even values of $n=2$ and $4$, it can be seen from (\ref{h1.8}) that the matrix $\tau_2$ plays no role in the angular boundary condition \cite{crespi1,crespi2}. This is due to the fact that for even $n$, $e^{i2\pi[-\frac{n\tau_2}{4}]} = \cos( {\frac{ \pm n \pi}{2}} ) = 
e^{i \pi [\frac{ \pm n }{2}]}$. For odd $n$, the two Dirac points are identified and the energy eigenstates of the Dirac equation are obtained by diagonalizing $\tau_2$ which mixes the valley indices in the wavefunction \cite{crespi1,crespi2}. Since the eigenvalues of $\tau_2$ are $\pm 1$, for all values of $n$, we can replace the angular boundary condition (\ref{h1.8}) by
\begin{eqnarray}
\label{h1.9}
\Psi(\mathbf{r},\theta=2\pi)=-e^{i2\pi[\pm \frac{n}{4}+(1-\frac{n}{6}) \frac{\sigma_3}{2}]}\Psi(\mathbf{r},\theta=0).
\end{eqnarray}
In subsequent analysis, we shall use the form (\ref{h1.9}) of the angular boundary condition assuming that for all allowed odd values of $n$, the matrix $\tau_2$ acting on the valley indices has already been diagonalized.

The effect of the angular boundary condition (\ref{h1.9}) on the wave function can be equivalently described by introducing a magnetic flux tube passing through the apex of the cone \cite{crespi1}. The presence of a magnetic vector potential modifies the boundary condition on a Dirac spinor as 
\begin{eqnarray}
\label{h5}
\Psi(\mathbf{r},\theta=2\pi)=-e^{ie\oint\vec{A}\cdot\vec{dl} }\Psi(\mathbf{r},\theta=0).
\end{eqnarray}
Taking $\vec{dl}$ as a line element on the circumference of the cone at a distance $r$ from the apex, we have
\begin{eqnarray}
\label{h6}
\vec{dl}= \hat{e}_{\theta}~r(1-\frac{n}{6})d\theta. 
\end{eqnarray}
Substituting (\ref{h6}) in (\ref{h5}) and assuming that $A_\theta$ is independent of the angle $\theta$, we get
\begin{eqnarray}
\label{h7}
\Psi(\mathbf{r},\theta=2\pi)=-e^{ie 2\pi r(1-\frac{n}{6})A_\theta}\Psi(\mathbf{r},\theta=0).
\end{eqnarray}
Comparing equation (\ref{h1.9}) and equation (\ref{h7}) we get
\begin{eqnarray}
\label{h8}
A_\theta = \frac{1}{er}[\pm\frac{\frac{n}{4}}{(1-\frac{n}{6})}+\frac{\sigma_3}{2}].
\end{eqnarray}

Thus the entire effect of the conical topology on the wavefunction can be described by introducing the vector potential (10), which replaces the ordinary derivatives by the corresponding covariant derivatives in the Hamiltonian (\ref{h1.1}). In this effective description, the Hamiltonian for a graphene cone with a Coulomb charge at the apex is given by
\begin{eqnarray}
\label{h9}
H=-i(\sigma_1 \partial_{x^\prime} + \sigma_2 \partial_{y^\prime})- e(\vec{\sigma}\cdot\hat{e}_{\theta}) A_{\theta} - \sigma_0 \left(\frac{-\alpha}{r}\right),
\end{eqnarray}
where the relation between $(x^\prime,y^\prime)$ and $(r,\theta)$ can be obtained from the Equation(\ref{h1.5}), which gives $\partial_{x^\prime} = \frac{1}{r(1-\frac{n}{6})}\partial_{\theta}$ and $\partial_{y^\prime} = -\partial_r$ and $\vec{\sigma}\cdot\hat{e}_{\theta}=\sigma_1$. 
Thus the Dirac equation (\ref{h1.1}) can be written as
\begin{eqnarray} 
\label{h11}
H \left( 
\begin{array}{c}
 \Psi_{A} \\
 \Psi_{B} 
\end{array}
\right) = \left( 
\begin{array}{cc}
-\frac{\alpha}{r} &  \partial_r - \frac{i}{r(1-\frac{n}{6})}\partial_\theta \pm \frac{\frac{n}{4}}{r(1-\frac{n}{6})} + \frac{1}{2r}  \\
-\partial_r - \frac{i}{r(1-\frac{n}{6})}\partial_\theta  \pm \frac{\frac{n}{4}}{r(1-\frac{n}{6})}-\frac{1}{2r} & 
 -\frac{\alpha}{r}
\end{array}
\right)
\left( 
\begin{array}{c}
 \Psi_{A} \\
 \Psi_{B} 
\end{array}
\right) = E \left( 
\begin{array}{c}
 \Psi_{A} \\
 \Psi_{B} 
\end{array}
\right)
\end{eqnarray}
We use an ansatz for the wavefunction given by 
\begin{eqnarray}
\label{h12}
\Psi(r,\theta)=\sum_j \left( 
\begin{array}{c}
\Psi_A^{(j)}(r)\\
i\Psi_B^{(j)}(r)
\end{array}
\right)e^{-iEr} {r}^{\gamma - \frac{1}{2}}e^{ij\theta},
\end{eqnarray}
where the total angular momentum $j$ takes all half integer values. Substituting 
(\ref{h12}) in (\ref{h11}), we note that the leading short distance behaviour of the wavefunction is given by 
\begin{equation}
\label{h16}
\Psi_{A,B}^{(j)}(r)\sim r^{\gamma-\frac{1}{2}}~~~~\mbox{where}~~~~\gamma = \sqrt{\nu^2 - \alpha^2}~~~~~~~\mbox{and}~~~~~~~\nu=\frac{(j\pm\frac{n}{4})}{(1-\frac{n}{6})}.
\end{equation}
From (\ref{h16}) it follows that when $|\alpha|$ exceeds $|\nu|$, $\gamma$ becomes imaginary. As a result, the eigenstates $\Psi_A^{(j)}(r)$ 
and $\Psi_B^{(j)}(r)$ becomes wildly oscillatory and have no well defined limit as $r\rightarrow0$, which corresponds to the phenomenon of zitterbewegung in a strong Coulomb electric field \cite{greiner}. The critical value of the coupling is denoted by $\alpha_c$ and it is given by the minimum allowed value of $|\nu|$. The parameter $\nu$ depends on $j$ and the number of sectors $n$ removed from a plane to form the graphene cone. Hence we see that the critical coupling $\alpha_c$ explicitly depends on the angle of the graphene cone. The values of the critical charge for different values of n are given in TABLE \ref{tab1}. 

\begin{table}
\caption{The values of critical charge $\alpha_c$ i.e the minimum values of $|\nu|$ for different values of opening angle of the 
         graphene cone i.e for different values of $n$. The values of angular momentum $j$ for which we get the critical charge are also 
         noted.}
\begin{tabular}{|c|c|c|}
\hline
value of $n$ & Critical charge $(\alpha_c)$ & Corresponding angular momentum $(j)$ \\
\hline
\hline
0 & $0.5$ & $\pm \frac{1}{2}$ \\
\hline
1 & 0.3 & $\pm \frac{1}{2}$  \\
\hline
2 & 0 &  $\pm \frac{1}{2}$ \\
\hline
3 & 0.5 &  $\pm \frac{1}{2}$  \\
\hline
4 & 1.5 &  $\pm \frac{1}{2}$ \\
\hline
5 & 1.5 &  $\pm \frac{3}{2}$  \\
\hline
\end{tabular}
\label{tab1}
\end{table}  

From these values of critical charge, a surprising result can be observed for the case of $n=2$, where $\alpha_c$ is zero. In this case, any external charge will be supercritical.We should also note that when $n=5$, none of the values of $\nu$ in the lowest angular 
momentum channel $j=\pm \frac{1}{2}$, corresponds to the minimum value of $|\nu|$ or the critical charge. In this case the critical charge
of the system is obtained from the angular momentum channel $j=\pm\frac{3}{2}$. 

\section{Spectrum in graphene cone with a supercritical charge}

In this Section we derive the scattering and bound state spectra for a gapless graphene cone in the presence of a supercritical Coulomb charge located at the apex of the cone. 
First consider the case $\nu \neq 0$. Using (\ref{h11}) and (\ref{h12}), the Dirac equation in each angular momentum channel $j$ can be written as
\begin{eqnarray}
\label{g2}
\left( 
\begin{array}{cc}
E+\frac{\alpha}{r} & -\{ \partial_r +(\nu +\frac{1}{2})\frac{1}{r}\}  \\
\{\partial_r -(\nu - \frac{1}{2})\frac{1}{r}\} & E+\frac{\alpha}{r}
\end{array}
\right) \left( 
\begin{array}{c}
\Psi_A^{(j)}(r)\\
i\Psi_B^{(j)}(r)
\end{array}
\right)e^{-iEr} {r}^{\gamma - \frac{1}{2}}=0.
\end{eqnarray}
In terms of the functions $u^{(j)}(r)$ and $v^{(j)}(r)$ defined by  $\Psi_A^{(j)}(r) = [v^{(j)}(r)+u^{(j)}(r)]$ and $\Psi_B^{(j)}(r) = [v^{(j)}(r)-u^{(j)}(r)]$, we get
\begin{equation} \label{g5}
r \frac{d v^{(j)}(r)}{d r} + (\gamma + i\alpha) v^{(j)}(r) - \nu u^{(j)}(r) = 0~~~~~~~~~
\end{equation}
and 
\begin{equation} \label{g6}
r \frac{d u^{(j)}(r)}{d r} + (\gamma - i \alpha -2iEr) u^{(j)}(r) - \nu v^{(j)}(r) = 0.~~~~~~~~~
\end{equation}
From (\ref{g5}) and (\ref{g6}), we get
\begin{equation} \label{gl7}
 s \frac{d^{2} v^{(j)}(s) }{d s^{2}} + (1 + 2\gamma - s)\frac{d v^{(j)}(s)}{d s} - \left ( \gamma + i \alpha \right ) v^{(j)}(s) = 0,
\end{equation}
where $s= -2ikr$, with  $k=- E$. For the discussion below, for any given value of $n$ and $j$, we choose $\alpha$ greater than the corresponding value of $|\nu|$. This ensures that the coupling is always in the supercritical region. Define $\gamma = i\lambda$ where $\lambda=\sqrt{\alpha^2 - \nu^2}$. Then the solution of 
Eq.(\ref{gl7}) is given by
\begin{equation} \label{gl8}
 v^{(j)}(s) = C_1 M \left ( i(\lambda + \alpha),~ 1 + 2i\lambda,~s \right)
    + C_2 {s}^{- 2 i\lambda} M \left (i(\alpha-\lambda),~ 1-2i\lambda,~s \right).
\end{equation}
From (\ref{g5}) and (\ref{gl8}) we get
\begin{equation} \label{gl10}
 u^{(j)}(s) = -iC_1 \mu M \left (1+i(\lambda + \alpha),~ 1 + 2i\lambda,~s \right)
    -i (C_2/\mu) {s}^{- 2 i\lambda} M \left (1+i(\alpha-\lambda),~ 1-2i\lambda,~s \right)
\end{equation}
where $\mu=\sqrt{\frac{\alpha+\lambda}{\alpha-\lambda}}$.

In order to proceed, we use the zigzag edge boundary condition $[u^{(j)}(a_0) - v^{(j)}(a_0)]= 0$, where $a_0$ is a distance from the apex, of the order of the lattice scale in graphene \cite{levi1}. This gives
\begin{eqnarray}
\label{gl13}
C_2 = e^{2i\zeta(k)} \mu e^{\pi \lambda} C_1~~~~~~\mbox{where}~~~~~e^{2i\zeta(k)} = \frac{i(1+i\mu)}{(1-i\mu)}e^{2i\lambda \mbox{ln}(2ka_0)}.
\end{eqnarray}
From the above, we obtain the scattering matrix $S$ as 
\begin{eqnarray}
\label{gl14}
S = e^{2i\delta_\nu (k)} =  \left [ \frac{f_{\alpha,\lambda}+e^{2i\zeta(k)}e^{-\pi \lambda}\mu f_{\alpha,-\lambda}}{e^{\pi \lambda}\mu f^{*}_{\alpha,-\lambda}+e^{2i\zeta(k)}f^{*}_{\alpha,\lambda}} \right ]
e^{-2i\alpha \mbox{ln}(2kr)}
\end{eqnarray}
where $f_{\alpha,\lambda}=\frac{\Gamma(1+2i\lambda)}{\Gamma(1+i\lambda-i\alpha)}$. From (\ref{gl14}) we obtain the scattering phase as 
\begin{eqnarray}
\label{gl15}
\delta_\nu (k) = \mbox{arg}[e^{-i\zeta(k)} + be^{i\zeta(k)}] - \alpha \mbox{ln}(2kr) + \mbox{arg}(f_{\alpha,\lambda})
\end{eqnarray}
where $b=e^{-\pi \lambda}\mu \frac{f_{\alpha,-\lambda}}{f_{\alpha,\lambda}}$. 
In Fig. (3a) we plot (\ref{gl15}) ignoring the Coulomb tail term $-\alpha ln(2kr)$. For $\nu=0.3$, the value of $\alpha=1.8$ is deeper in the supercritical region than it is for $\nu = 1.5$. When the coupling $\alpha$ is deeper in the supercritical region, the phase shift has more number of kinks, which indicate the bound states. The plot also shows that the phase shift depends on the topology through its dependence on $n$ via $\nu$.
\begin{figure}
\centering
\begin{tabular}{cc}
\epsfig{file=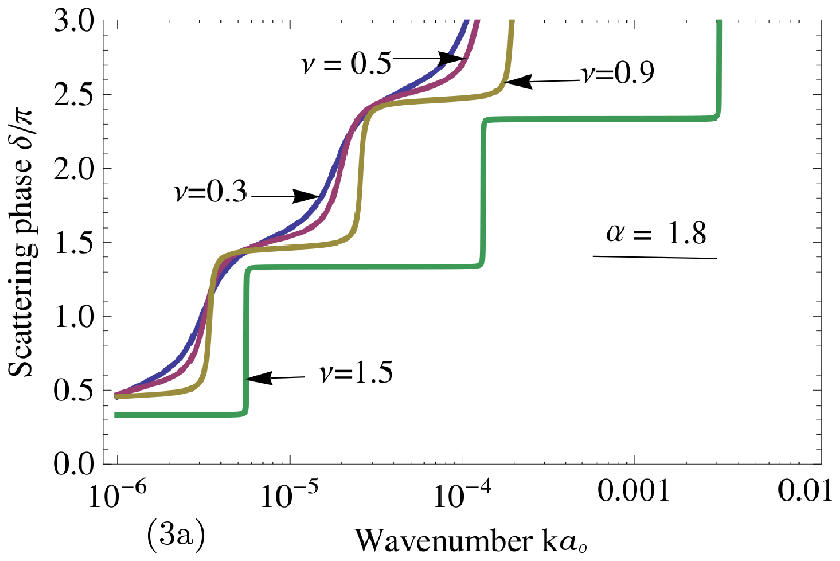,width=0.5\linewidth,clip=} &
\epsfig{file=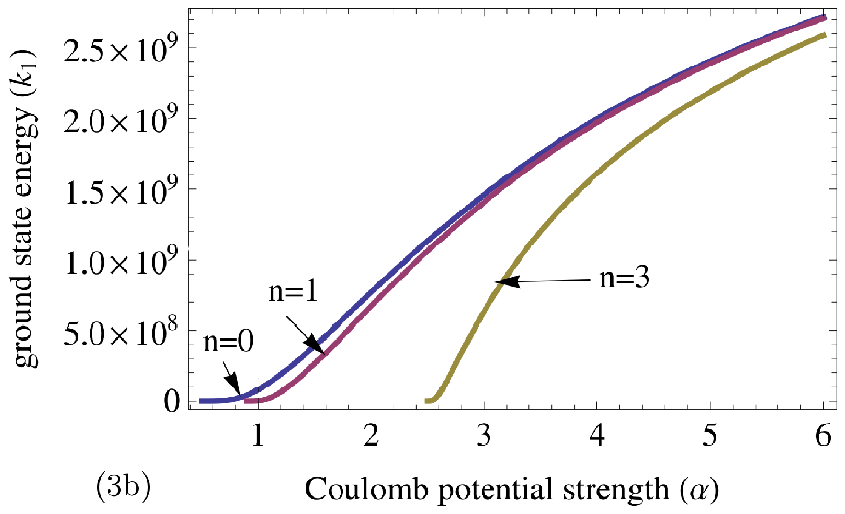,width=0.5\linewidth,clip=}\\
\end{tabular}
\caption{(3a) shows the dependence of scattering phase shift $\delta$ on wavenumber $ka_0$ for $\nu = 0.3, 0.5, 0.9, 1.5$ and $\alpha=1.8$, 
ignoring the Coulomb tail term $-\alpha ln(2kr)$. As the value of $\nu$ increases, the kinks in the phase shift become sharper, which indicates 
the dependence of the phase shift on the angle of the graphene cone. (3b) shows dependence of ground state energy on the Coulomb potential 
strength for different angles of the graphene cone. We have considered $\nu=\frac{j+\frac{n}{4}}{1-\frac{n}{6}}$ and $j=\frac{1}{2}$.}
\label{fig:4}
\end{figure}

As mentioned before, in gapless graphene we do not expect bound states due to Klein tunneling. However, in the supercritical regime, the 
system admits quasi-bound states whose spectrum is obtained from the zeroes of the $S$ matrix in (\ref{gl14}). The quasi-bound state energies 
are given by
\begin{eqnarray}
\label{gl16}
k_p = \frac{1}{2a_0}\mbox{exp}\left[-\frac{p\pi}{\lambda} +i \left(\frac{1}{2\alpha}-\frac{\pi}{2}\right)\right],
\end{eqnarray} 
where $p$ is a positive integer. 

\begin{figure}
\centering
\begin{tabular}{cc}
\epsfig{file=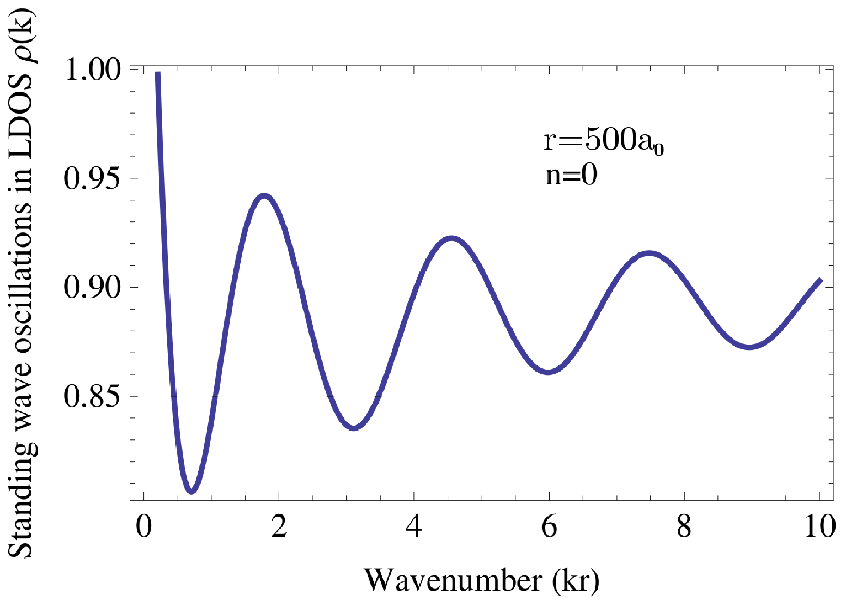,width=0.5\linewidth,clip=} &
\epsfig{file=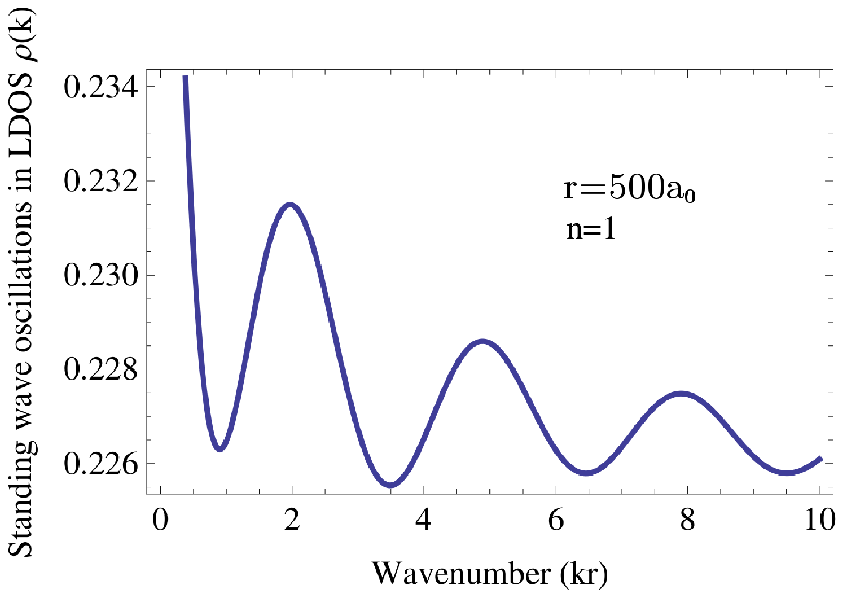,width=0.5\linewidth,clip=}\\
\end{tabular}
\caption{Energy dependence of LDOS in presence of a Coulomb potential for different values of $n$ and a particular value of $r$ and with $j=\frac{1}{2}$ and $\alpha =0.6$. This is in the supercritical region for both $n=0$ and $1$. }
\label{fig:6}
\end{figure}

Another interesting observable in this context is the LDOS. In Fig.(\ref{fig:6}) we have plotted the standing wave oscillations in LDOS $\rho(k,r)$ using 
\begin{eqnarray}
\label{ldos}
\rho(k,r)=\frac{4}{\pi \hbar v_F}\sum_{j}|\Psi^{(j)}(k,r)|^2,
\end{eqnarray}
where $\Psi^{(j)}(k,r)$ is the radial part of the spinor given in Eq.(\ref{h12}), for a given angular momentum channel $j$. 
The maxima and minima of the oscillations occur at half-integral and integral multiple of $\pi$ respectively. The normalization constant of the wave function depends on the value of $n$ which decreases with increasing $n$. Therefore, though the nature of standing wave oscillation is similar for $n=0$ and $1$, the amplitudes of the corresponding LDOS decreases with the increase of $n$.  

It is known that in the planar case, under the RG flow, a supercritical charge is driven to its critical value \cite{levi1}. While the full discussion of the RG flow requires a more detailed analysis, we can obtain some of its qualitative features from the quasi-bound state energies. To see this consider the real part of $k_p$ in (\ref{gl16}) which gives the energy. This part diverges as the cutoff $a_0 \rightarrow 0$. In order to study the RG flow, we now promote the coupling constant $\alpha$ as a function of $a_0$ and demand that as $a_0 \rightarrow 0$, the energy for any fixed level $p$ (say $p = 1$) remains independent of the cutoff \cite{rajeev,us2}. In the leading order, where $\alpha$ is only slightly above the critical coupling, this prescription gives the $\beta$-function as 
\begin{equation}
\label{beta}
\beta (\lambda) \sim - \lambda^2 + ..
\end{equation}
Thus we have an ultraviolet stable fixed point at $\lambda = 0$ or at $\alpha = \nu$. Hence, for any given value of $n$ and $j$, the coupling $\alpha$ in the supercritical regime is driven to its critical value.

We end this Section with a comment for the case $\nu=0$, which occurs for $n=2$ and $j=\pm \frac{1}{2}$. For $\nu=0$, the first order Dirac equations (\ref{g5}) and (\ref{g6}) completely decouple and the functions $u$ and $v$ become independent of each other. In this case, the critical coupling $\alpha_c$ also vanishes. This means that any external charge, no matter how small, would lead to the supercritical regime.

\section{Conclusion}

In this paper we use the fact that system of a graphene cone with an external Coulomb charge at its apex can be equivalently described by the combination of the Coulomb charge and a suitable magnetic flux tube passing through the apex. The above analysis has been done in the supercritical regime, where the external Coulomb charge exceeds a certain critical value. 

The quantities of physical interest in this system include the scattering phase shifts, the LDOS and the quasi-bound state energies. We have shown that all these physical quantities depend explicitly on the number of sectors removed from a planar graphene to form the cone. The existence of the quasi-bound states indicates the possibility of the localization of the wavefunctions in the presence of a supercritical charge. Our analysis shows that the nature and extent of the localization depends on the spatial topology of the graphene sample.

Finally, we have given qualitative arguments which shows that under the RG flow and for $\nu \neq 0$, the supercritical charge in the graphene cone tends to its critical value. If this argument can be extended for $\nu=0$, for which the critical charge vanishes, that would lead to complete shielding of the external charge. This issue and the related electronic properties are currently under investigation.

\end{document}